\documentclass[prb,preprint]{revtex4-1} 

\usepackage{amsmath}  
\usepackage{amsfonts} 
\usepackage{graphicx} 
\usepackage{ulem}

\usepackage[hidelinks]{hyperref}

\usepackage[dvipsnames]{xcolor}

\newcommand{\name}{Nuclear Beavers }
\newcommand{\institution}{the Facility for Rare Isotope Beams}

\begin{document}


\title{Nuclear Beavers}

\author{J. Wylie}
\affiliation{Facility for Rare Isotope Beams, Michigan State University, East Lansing, Michigan 48824, USA}
\affiliation{Department of Physics and Astronomy, Michigan State University, East Lansing, Michigan 48824, USA}

\author{P. Giuliani}
\affiliation{Facility for Rare Isotope Beams, Michigan State University, East Lansing, Michigan 48824, USA}
\affiliation{Department of Physics and Astronomy, Michigan State University, East Lansing, Michigan 48824, USA}

\author{S. Agbemava}
\affiliation{Facility for Rare Isotope Beams, Michigan State University, East Lansing, Michigan 48824, USA}

\author{K. Godbey}
\affiliation{Facility for Rare Isotope Beams, Michigan State University, East Lansing, Michigan 48824, USA}
\affiliation{Department of Physics and Astronomy, Michigan State University, East Lansing, Michigan 48824, USA}

\vfill
\begin{center}
\footnotesize
The following article has been submitted to/accepted by \textit{The Physics Teacher}. 
After it is published, it will be found at 
\href{https://doi.org/10.1119/5.0253360}{https://doi.org/10.1119/5.0253360}.
\end{center}
\vspace*{2em} 

\maketitle

Atomic nuclei are the intricate cores that sit at the heart of atoms, the building blocks of matter.
The nucleus itself is made from protons and neutrons, which we call nucleons. 
The number of protons will determine the specific element, such as carbon, oxygen, or gold, while the number of neutrons will determine variations of those elements, which we call isotopes.
Some combinations of protons and neutrons last forever, while others are radioactive and can decay.
Far from being fully understood, atomic nuclei continue to be an active field of study in both theoretical and experimental science. 
Nuclei have significant implications not only for basic scientific inquiry, but also for various practical applications ranging from energy production to medical imaging\cite{conversation_radioactive_tracers}.
The nuclear landscape presents a tapestry of wonders: shapes and deformations, peculiar decays and transmutations, reactions and dynamics, and complex structures all coalesce to give each nucleus its own spotlight in the scientific arena.

\begin{figure}[!ht]
    \centering
    \includegraphics[width=0.7\linewidth]{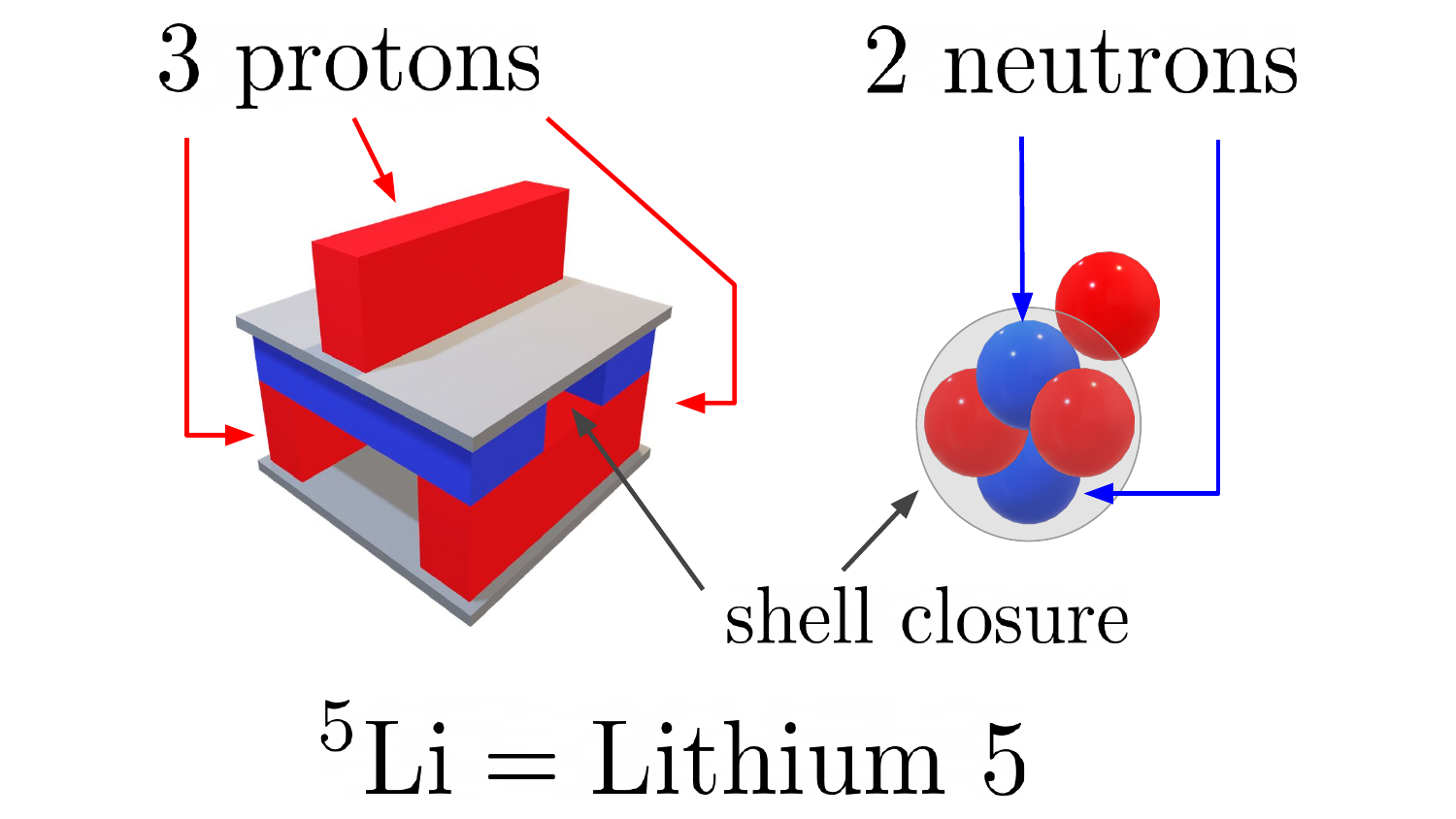}
    \caption{Representation of lithium-5 ($^5$Li). The element symbol, in this case $\text{Li}$, is complemented by the superscript number that accounts for the total number of nucleons, in this case 5. Right: we show the traditional view showcasing protons and neutrons as red and blue spheres, respectively. Left: we show the \name approach where protons and neutrons are represented as red and blue blocks, respectively. The gray platform and gray transparent bubble are used to represent a shell closure, which encapsulates a group of nucleons that are bound tightly together.}
    \label{fig:schematics}
\end{figure}

The Nuclear Beavers demonstration provides a hands-on experience that offers an intuitive lens into nuclear structure and decay.
Beavers are known for their capacity to build elaborate structures from wood, and this activity uses wooden blocks to allow participants to engage with the complexities of atomic nuclei.
To tie \name to reality, we show in Figure~\ref{fig:schematics} the structure of the lithium-5 nucleus.
We aim to provide a more accessible entry point for students and educators alike by converting complex nuclear interactions and structures into tactile ``building blocks" with defined rules, allowing everyone to engage with this fascinating part of science.

\section*{Characteristics of nuclear physics}

Bounded by the strong nuclear force, one of the four fundamental forces in nature, the protons and neutrons inside a nucleus form a rich spectrum of possibilities.
From hydrogen (the lightest element) to oganesson (the heaviest element), there are numerous scientific descriptions of these nuclei with different focuses to tackle specific problems.
Similarly, different pedagogical approaches for nuclear science have been provided to address some of these nuclear features.
To help students understand important concepts, we first discuss other demonstrations in the field of nuclear physics before detailing our new approach.

An intuitive method for explaining the collisions of nuclei in laboratories, for example at \institution, was developed by Constan~\cite{Constan2010}.
This showcases the possible products from two nuclei impacting against each other but does not highlight any significant nuclear structure effects.
Whittaker \cite{Whittaker2013} created a method to highlight how quarks make up protons and neutrons which in turn form larger nuclear systems.
This illustrates how complex behaviors like magic numbers emerge naturally from different combinations of protons and neutrons, yet only a static picture of these nuclei are accessible in the demonstration.

Although both models highlight very important aspects of nuclear physics, they focus on two extremes of nuclear observation.
Whittaker's approach starts from the smallest scales to create static nuclei to highlight their structure, while Constan's approach illustrates the relation between the initial and final products of reactions.
However, nuclei evolve over time and some constructions that might appear stable will inevitably decay after a certain time has passed due to quantum mechanical effects.
The \name demonstration we introduce complements both preceding papers as it allows for the construction of nuclei, while introducing vibrations (mimicking quantum mechanics in the real world) and excitations to break these nuclei over time.

The two main concepts \name highlights are the arrangement of protons and neutrons within the nucleus in shell-like structures (the shell model~\cite{conversation_magic_numbers}) and the fact that most nuclei are unstable and undergo decay.
We focus on three types of decays: gamma($\gamma$)-decay, beta($\beta^\pm$)-decay, and nucleon emission.
The physics of these phenomena are elaborated in this manuscript, while the detailed descriptions on how to build the demonstration along with the rules can be found in the Supplemental Material~\cite{Sup}.
This demonstration can also be used in a general lecture-type environment to illustrate the complexities of nuclear physics to a class.

\section*{Concepts}\label{sec:structure}

\subsection*{Electric charge}

This demonstration helps students engage with specific challenges in nuclear formation, particularly the interplay between the Coulomb and strong nuclear force.
The Coulomb force, or force due to electric charge, is noticeable for protons but not for neutrons. Both nucleon species experience the strong nuclear force which binds nucleons together to make a nucleus.
When two objects with the same charge approach each other, they repel, just like other electric charge demonstrations.
Therefore, protons prefer to avoid being near each other to minimize the repelling force they create to avoid breaking the nucleus apart.
This tension is alleviated by adding extra neutrons which also bind through the strong nuclear force but without the Coulomb repulsion.

Looking at the light nuclei, many of the stable systems have equal numbers of protons ($Z$) and neutrons ($N$), but as we move towards heavier systems more neutrons are needed to maintain stability.
We illustrate this in a companion website~\cite{Wylie2024} where one can select the ``$N=Z$" chart option to draw a line highlighting this deviation.

Students can further explore this phenomenon after being introduced to the rules~\cite{Sup}.
To explore the effects of charge, we suggest players try to build nuclei beyond $^{40}$Ca if possible, both following and not following the proton spacing rule.
Not following the rule could lead participants to build very large and stable nuclei with equal numbers of protons and neutrons, contrary to nature.
This highlights that in science, the discrepancies between models and data can elucidate more fundamental understanding.
In our model, if we do not include charge, we notice experimental data is not reproduced, driving a need to improve the model by including charge.
This process is commonly referred to as the scientific method.

\subsection*{Shell structure}

\begin{figure*}[!h]
    \centering
    \includegraphics[width=\linewidth]{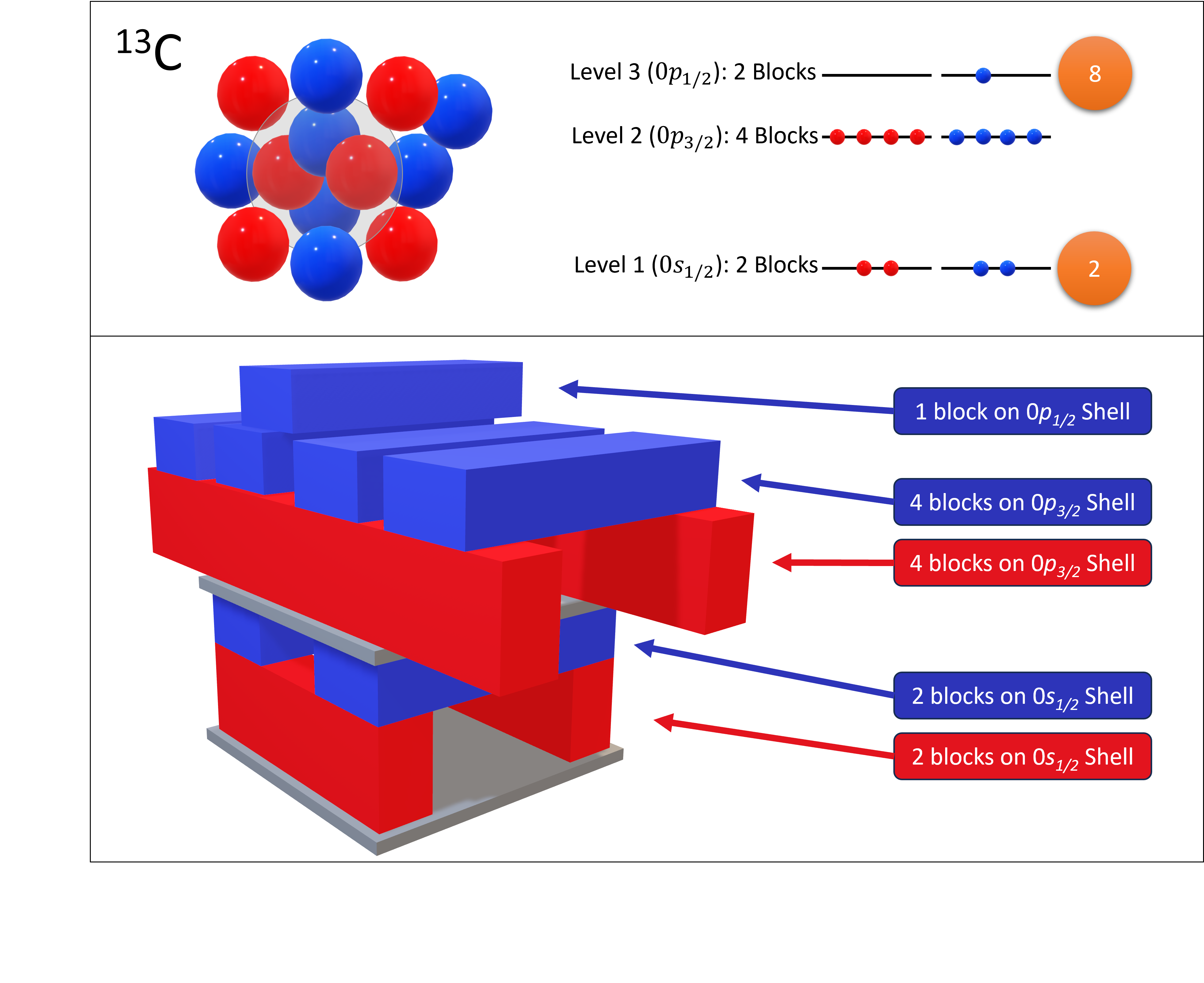}
    \caption{Top: Shell model picture of carbon-13 ($^{13}$C) with filled $0s_{1/2}$ shell (magic number at 2), filled $0p_{3/2}$ shell, and partially filled $0p_{1/2}$ shell. The number of nucleons which fit in the shell are given while the prefix number e.g. 0 in $0s_{1/2}$ indicates the first instance of a shell of such type (subsequent shells of the same type would have a prefix of 1, 2, etc.). Bottom: illustration of $^{13}$C using the block structure. In both cases, protons and neutrons are represented as red and blue respectively.}
    \label{fig:shell_illustration_physics_only_paper}
\end{figure*}

Nuclear structure is complex due to the numerous ways nucleons can orient themselves~\cite{conversation_magic_numbers}.
One of the first methods developed to describe nuclear structure is the nuclear shell model, which assumes nucleons prefer to construct a nucleus in a certain order.
The ordering is referred to as a ``shell" and, in the case of our demonstration, these are the levels of blocks we build, see Figure~\ref{fig:shell_illustration_physics_only_paper} for an example of shells.

Various shells can hold different numbers of nucleons and nuclear physicists refer to them as $s_{1/2}$, $p_{3/2}$, $d_{5/2}$, etc.
The letter of the shell, for example $s, p$, and $d$, relates to the angular momentum or how the nucleons orbit around the nucleus, analogous to how the earth rotates around the sun each year.
Nucleons can also spin or rotate about themselves which is analogous to how the earth rotates every 24 hours.
The subscript, for example $1/2$, $3/2$, and $5/2$, is the combination of the spin and angular momentum. 
We can calculate how many nucleons a level can accommodate by multiplying the subscript ($j$) by two and adding $+1$ for a total number $N_{total}=2j+1$. For example $0s_{1/2}$ and $0p_{3/2}$ contain 2 and 4 (calculated by $\left(\frac{1}{2}\times 2\right)+1=2$ and $\left(\frac{3}{2}\times 2\right)+1=4$) nucleons respectively, shown in Figure~\ref{fig:shell_illustration_physics_only_paper}.

Specific groups of shells, like $0s_{1/2}$ or the $0p_{3/2}$ and $0p_{1/2}$ group in Figure \ref{fig:shell_illustration_physics_only_paper}, enhance nuclear stability when completely filled.
The particular total numbers of protons and neutrons that produce this enhanced stability are referred to as “magic numbers”.
Magic numbers emerge as a result of increased spacing between subsequent shells.
In our demonstration, magic numbers correspond to the cardboard levels that increase stability.
These phenomena are highlighted with rules 4 and 5~\cite{Sup} as well as the companion website~\cite{Wylie2024}, where the bands underlying the chart correspond to each magic number.
A similar process occurs in electron structure in atoms where a specific number of electrons in a well-bound orbit produce non-reactive noble gases, such as helium, neon, or argon.

Nuclear shells can explain the presence of excited states.
Many nuclei contain excited states which are nuclear configurations where a nucleon is placed in a higher shell than its normal (ground) state.
These excited states are higher in energy, since it takes energy to promote a nucleon up, and states with higher energy tend to be more likely to decay.

\subsection*{Decays}\label{sec:decays}

\begin{figure*}[!h]
    \centering
    \includegraphics[width=\linewidth]{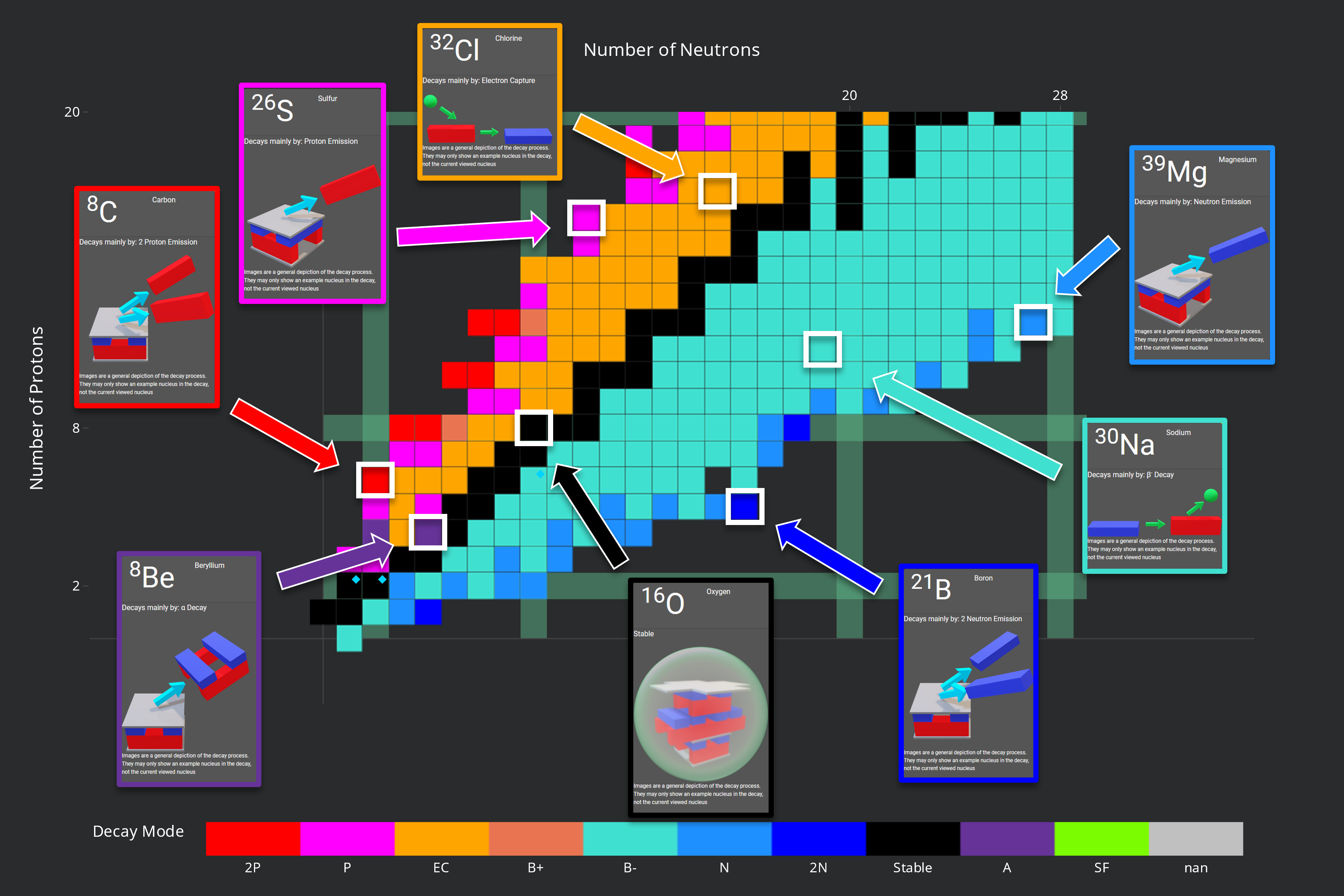}
    \caption{Nuclear landscape of all known nuclei up to $^{48}$Ca shown with examples of different decay modes from companion website~\cite{Wylie2024}. We note that users can submit any new structures they build in this demonstration to be highlighted as ``discoveries" within this site.}
    \label{fig:nuclear_landscape}
\end{figure*}

In this demonstration, towers resistant to vibration are analogous to stable nuclei whereas towers that undergo change due to vibration are analogous to unstable nuclei.
By following the level counting chart provided~\cite{Sup}, one can construct nuclei which tend to be more stable.
Notice these stable nuclei are located along the black squares in Figure~\ref{fig:nuclear_landscape} or the Valley of Stability~\cite{Wylie2024}, meaning that they are stable and more likely to exist than their neighboring isotopes.
Beyond this region, we find exotic nuclei with significantly shorter lifetimes, meaning they tend to decay faster or more easily fall apart in the demonstration.
In our demonstration, a longer lifetime would correspond to being able to withstand the table vibrations longer than other nuclei.

By not filling certain levels or creating an excess of one type of nucleon, we explore different phenomena like $\beta$ decay, $\gamma$ emission, or nucleon emission.
Fission is not explored within this set-up as it occurs in heavy nuclei which can be difficult to make as they require over 200 blocks.

\begin{figure}[!h]
    \centering
    \includegraphics[width=\linewidth]{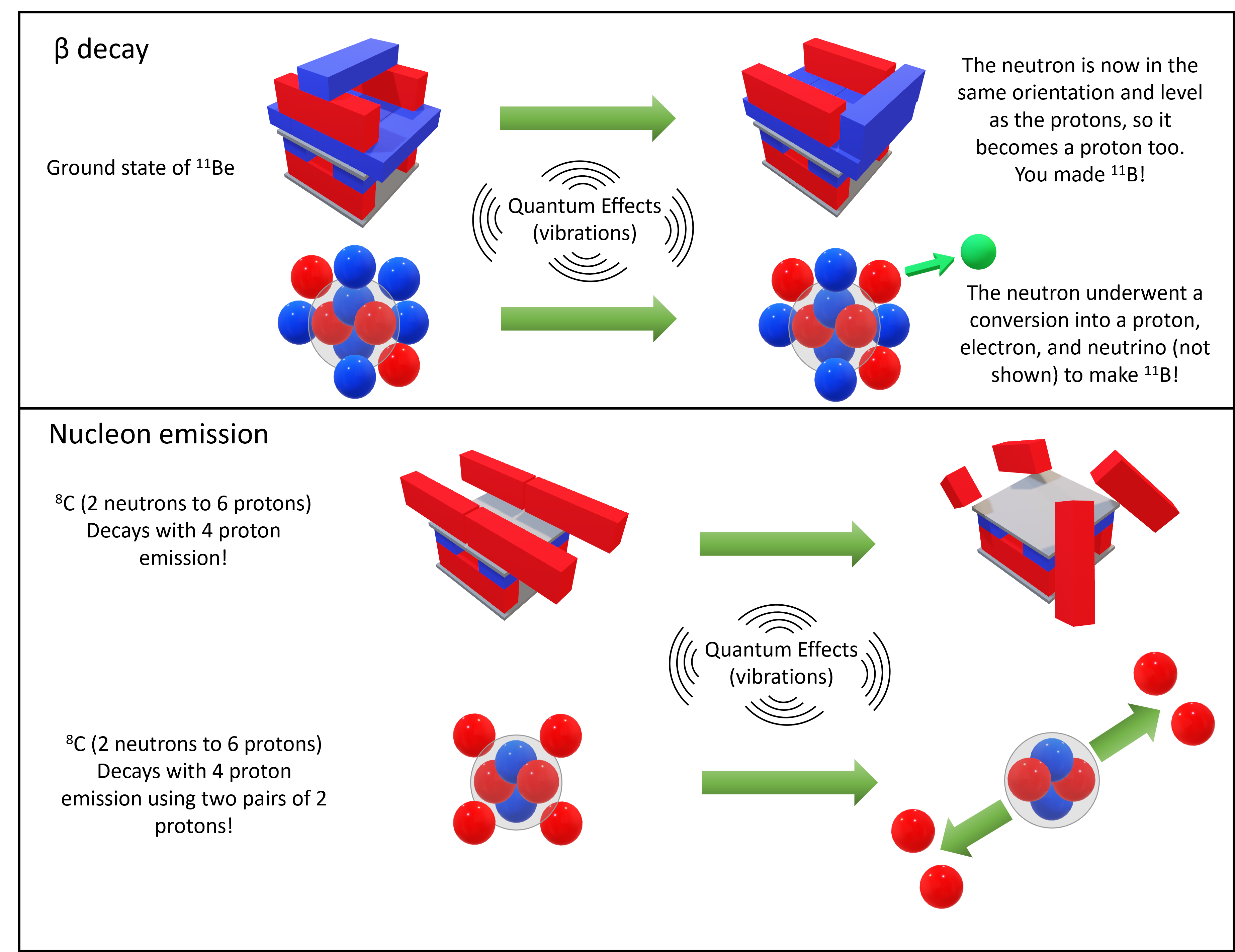}
    \caption{Two modes of decay are presented, but all three illustrated in the Supplemental Meterial~\cite{Sup}. Top: $\beta^- -$decay illustrating the conversion of a neutron to a proton. The \name representation does not include electrons but it is shown for charge considerations. The Nuclear Beavers representation does not include electrons or neutrinos but the electron is shown for charge considerations. Bottom: Nucleon emission example of $^8$C which decays by emitting four protons in two pairs of two protons.}
    \label{fig:decay_examples}
\end{figure}

A common nucleon decay mode across the nuclear chart is $\beta^\pm$-decay which plays an essential role in the creation of heavier nuclei~\cite{Spyrou2024}. There are actually two methods of $\beta$-decay, $\beta^+$ and $\beta^-$, which involve the conversion of a proton to neutron and neutron to proton respectively.
Neutrinos and antineutrinos will also be emitted in this process, but in this simplified model they are ignored.
Depending on the excess amount of one nucleon type, we find that $\beta$-decay favors the conversion of excess nucleons as lighter nuclei tend to prefer more balanced numbers of nucleons.
In our demonstration, we find that $\beta$-decay can emerge naturally, as one block might fall to another level or to its opposite orientation and convert into the other species.
For example, when we apply vibrations, if a blue neutron block falls onto its side so it stands like a red proton block or falls onto a proton level, it has now $\beta^-$-decayed into a proton, see Figure~\ref{fig:decay_examples}.
Conversely, when a red proton block falls and lands in the neutron orientation or neutron level, we can say that it has $\beta^+$-decayed into a neutron.

Another very common decay in nuclei is $\gamma$-decay, which does not involve a change in the number of protons or neutrons in a nucleus; instead, it removes energy from the nucleus in the form of light emission, the emission of a photon.
Being in an excited state is usually unfavorable because nuclei prefer to remain in their lowest energy state, so they choose to get rid of excess energy by emitting a photon to carry off the energy it wishes to discard.
$\gamma-$decay occurs in \name as blocks can fall to lower levels~\cite{Sup}.
One may expect to see more $\gamma-$decays while exciting nucleons to higher levels due to increased instability when adding more energy to move the nucleons.

The final decay mode we focus on is nucleon emission, where a certain number of nucleons are emitted simultaneously or sequentially.
Some nuclei decay primarily through one, two, or even more nucleon emission depending on how exotic they are, and more are still being discovered~\cite{Schirber2023}.
Beryllium-6 ($^6$Be) decays by emitting two protons simultaneously while carbon-8 ($^8$C) is known to decay by emitting four protons in pairs of two protons, shown in Figure~\ref{fig:decay_examples}.
This method of decay will likely be seen more frequently in this demonstration than in nature, because any particles falling off of the nucleus are classified as nucleon emission.

\section*{Conclusions}\label{sec:conclusions}

Nuclear physics can be difficult to include in course curriculum because of its abstract nature.
We encourage teachers to provide this demonstration to their students to facilitate exploration in both the structure of nuclei and their decays.
The goal is to let students explore on their own and in groups to create the largest and/or most exotic nuclei they can\cite{Sup}.
Through this hands-on activity, students will hopefully be able to identify: why nuclei favor more neutrons to protons, how excited states make it harder for nuclei to remain stable, and in what circumstances we expect to see different types of decays. Their answers may vary, but we hope they contain some or all of the following elements:
\begin{itemize}
    \item Protons repel each other like magnets because of their charge. Proton-rich configurations make the nucleus more unstable due to excess charge.
    \item Excited states made the bases of our towers more unstable by leaving holes in them.
    \item Excited nucleons are more likely to fall over since they're stacked on top of one another in levels.
    \item We see more decays when we excite more nucleons.
    \item Depending on how long we vibrate the table some nuclei remain stable but if we vibrate them for longer times they also eventually decay, so each nucleus has a different lifetime based on its stability.
\end{itemize}
We encourage students to use the companion website~\cite{Wylie2024} and (with proper permission) submit their own nuclear discoveries.

\section*{Acknowledgements}\label{sec:acknowledgements}

We thank all of our play-testers who provided feedback on areas to improve this demonstration. Special thanks to Zach Constan and \institution ~for providing materials and opportunities to present this to the public. Supported by the U.S. Department of Energy, Office of Science, Office of Nuclear Physics under award DE-SC0013365.

\setcitestyle{numbers}
\bibliographystyle{unsrt} 

\end{document}



\title{Supplemental Material for ``Nuclear Beavers"}

\author{J. Wylie}
\affiliation{Facility for Rare Isotope Beams, Michigan State University, East Lansing, Michigan 48824, USA}
\affiliation{Department of Physics and Astronomy, Michigan State University, East Lansing, Michigan 48824, USA}

\author{P. Giuliani}
\affiliation{Facility for Rare Isotope Beams, Michigan State University, East Lansing, Michigan 48824, USA}
\affiliation{Department of Physics and Astronomy, Michigan State University, East Lansing, Michigan 48824, USA}

\author{S. Agbemava}
\affiliation{Facility for Rare Isotope Beams, Michigan State University, East Lansing, Michigan 48824, USA}

\author{K. Godbey}
\affiliation{Facility for Rare Isotope Beams, Michigan State University, East Lansing, Michigan 48824, USA}
\affiliation{Department of Physics and Astronomy, Michigan State University, East Lansing, Michigan 48824, USA}

\maketitle

\section*{Materials}\label{sec:materials}

\begin{figure}[!h]
    \centering
    \includegraphics[width=\linewidth]{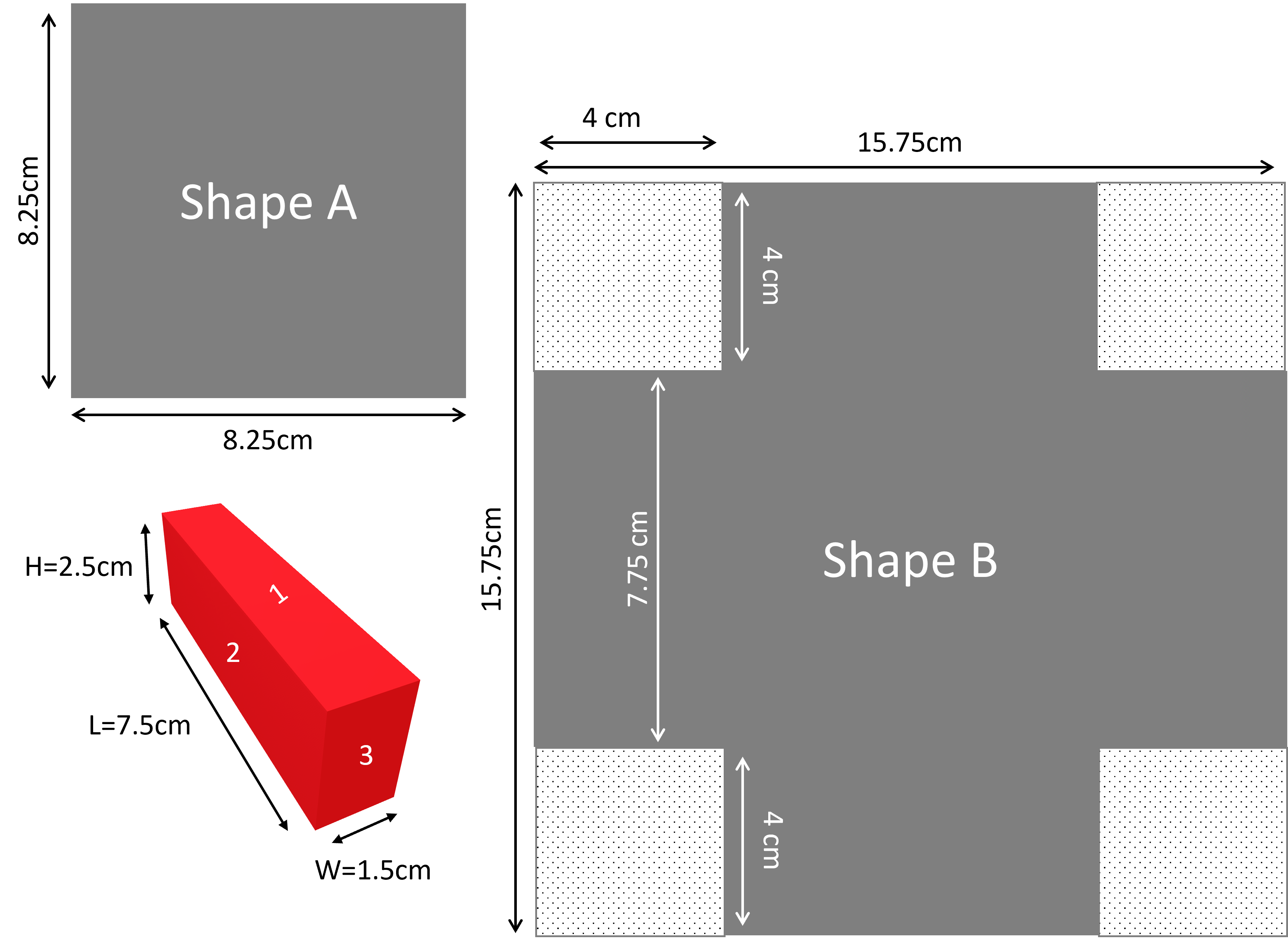}
    \caption{Diagram of blocks, cut cardboard pieces, and their dimensions. Shape A represents the starting base and first nuclear magic number (2). Shape B corresponds to the second and third nuclear magic numbers (8 and 20). The dotted white squares inside Shape B indicate the segments one must cut out of the 15.75cm x 15.75cm cardboard square. An example block with labeled sides and dimensions is shown for clarity in the rules.}
    \label{fig:shell_closures}
\end{figure}

Items required are wooden blocks with dimensions: 1.5cm × 2.5cm × 7.5cm and cardboard shapes which will act as extra support.
We recommend at least 48 wooden blocks with 20 painted red (protons) and 28 painted blue (neutrons).
The cardboard should be cut into specific shapes (dimensions are indicated in Figure \ref{fig:shell_closures}) and will serve as ``Magic Number shell closures" (see Concepts for details).

\section*{Rules}\label{sec:rules}

\begin{figure*}[!htb]
    \centering
    \includegraphics[width=\linewidth]{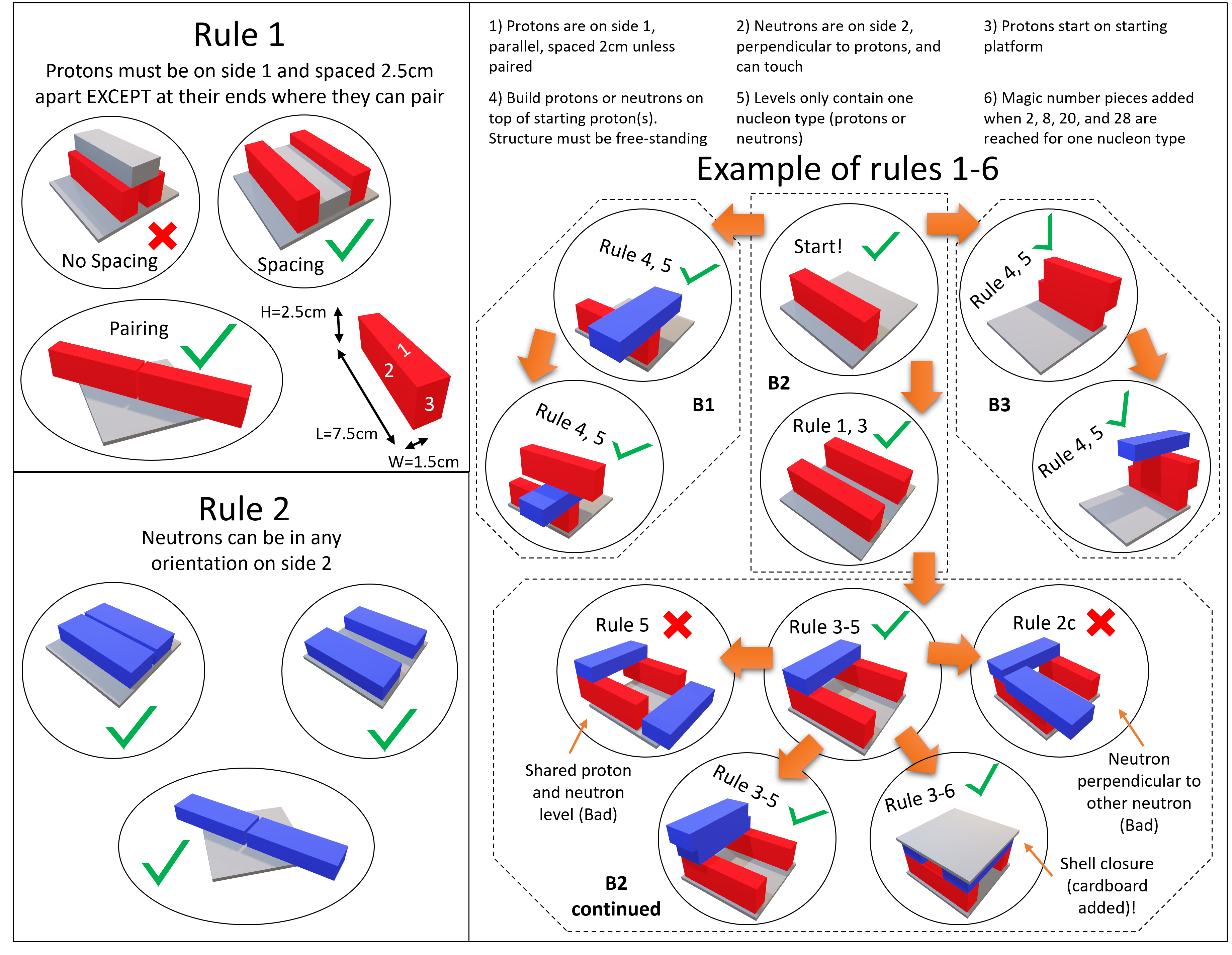}
    \caption{Illustrations of allowed (green check mark) and not allowed (red `x') block placements when constructing nuclei. We show a few possible starting moves in ``Examples of rules 1-6," to create helium-3 ($^3$He) and helium-4 ($^4$He) in B1/B3 and B2 respectively.}
    \label{fig:paper_rules}
\end{figure*}

Participants can play the demonstration individually or competitively with each other. The following rules are important for both versions, but first, we review terms for the demonstration.
Red blocks (protons) and blue blocks (neutrons) are your ``nucleons" with which you build your nucleus.
To begin any demonstration, place a proton onto the starting platform (shape A of Figure \ref{fig:shell_closures}).
This is the first ``level" or ``shell" of your chosen nucleon, and it must be supported only by the starting platform.
We build our nucleus by adding more blocks and levels to it.
The entire nucleus must be self-supported on the starting platform (no external support can be added).
Additional rules are as follows and an example is shown in Figure \ref{fig:paper_rules}:
\begin{enumerate}
    \item Protons:
    \begin{enumerate}
        \item Must be laying on side 1 (shown in Figure \ref{fig:shell_closures})
        \item Must be spaced $\approx$2cm apart (about the height of side 3)
        \item Must be parallel to one another
        \item Can ``pair" by having their side 3 touch
    \end{enumerate}
    \item Neutrons:
    \begin{enumerate}
        \item Must be laying on side 2 (shown in Figure \ref{fig:shell_closures})
        \item Can touch in any configuration
        \item Must be parallel to one another AND perpendicular to protons
    \end{enumerate}
    \item To start, at least one proton must be placed first on the starting platform (shape A in Figure \ref{fig:shell_closures})
    \item Subsequent protons or neutrons can be built on top of this starting level
    \item Each level can ONLY contain protons or neutrons (the order of proton or neutron levels doesn't matter after the start as long as no more than two levels of the same nucleon type are on top of each other)
    \item When a magic number is reached (2, 8, 20, 28 of one type of nucleon), you may place a piece of cardboard on top to help stabilize the structure
    \item (Optional) Each level of protons or neutrons has a maximum number of blocks allowed and you cannot place more than indicated in the current level
\end{enumerate}

\begin{figure*}[!htb]
    \centering
    \includegraphics[width=\linewidth]{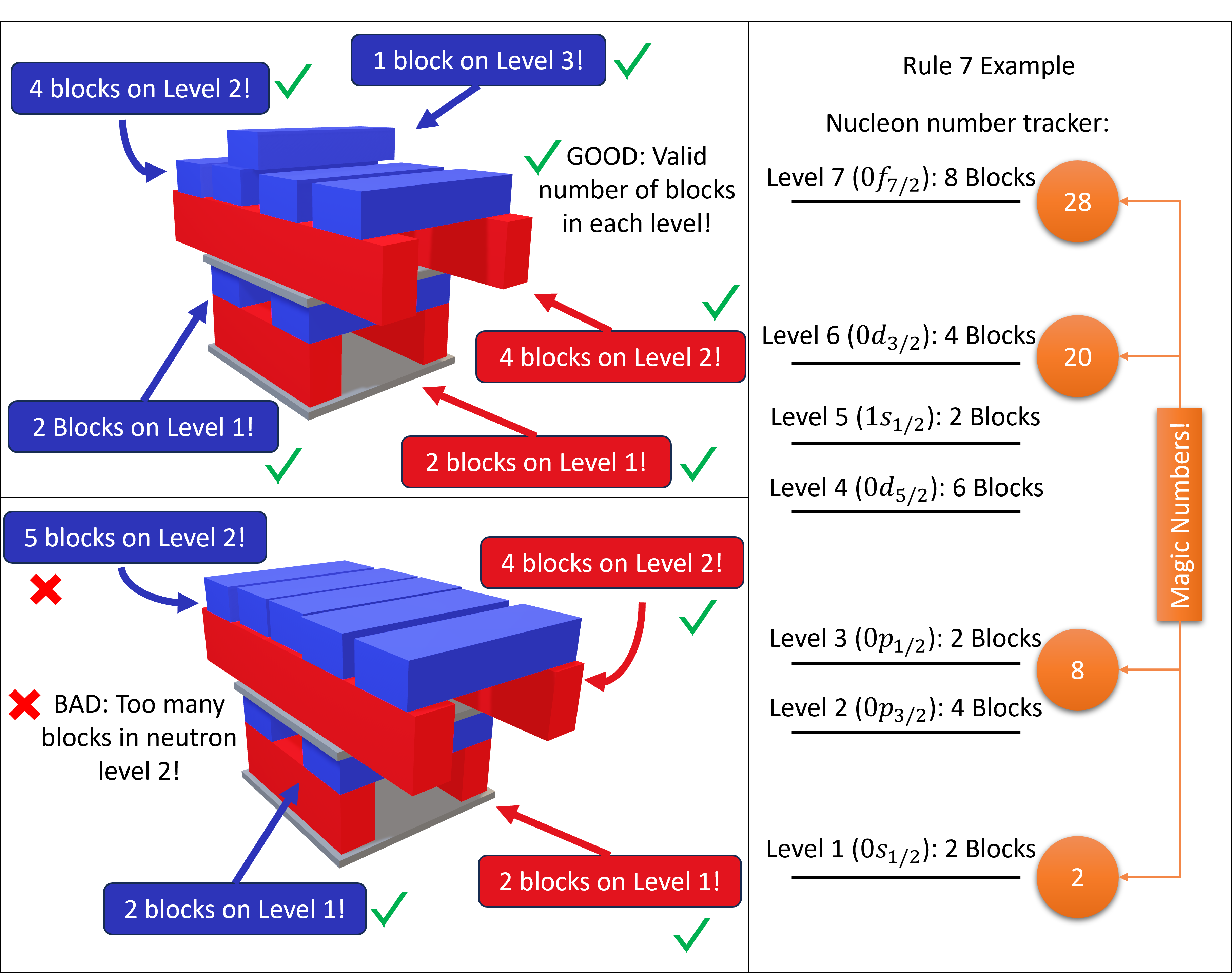}
    \caption{Illustrations of allowed (green check mark) and not allowed (red `x') block placements when following Rule 7. Shown also is the level diagram of usual orderings and their scientific name.}
    \label{fig:rule_7_and_level_diagram}
\end{figure*}

We note the impacts of electric charge, or lack thereof, on nuclei is taken into account by the first two rules.
There are two ways to build nuclei depending on the preference of the participants.
The first method is simpler and recommended for younger participants as it only follows the main 6 rules (shown in Figure \ref{fig:paper_rules}).
Rule 7 can give a deeper understanding of the physics involved in nuclei as the levels themselves are important.
These levels give us interesting shapes and sometimes can mix together or change in order to make more exotic nuclei.
The most realistic reproduction of nuclear structure occurs when using all 7 rules, which is encouraged for high school students and beyond.
If including rule 7, the levels provided in Figure~\ref{fig:rule_7_and_level_diagram} do not necessarily need to go in order, but the constructed nucleus is usually more stable when following this order. 

\begin{figure}[!h]
    \centering
    \includegraphics[width=\linewidth]{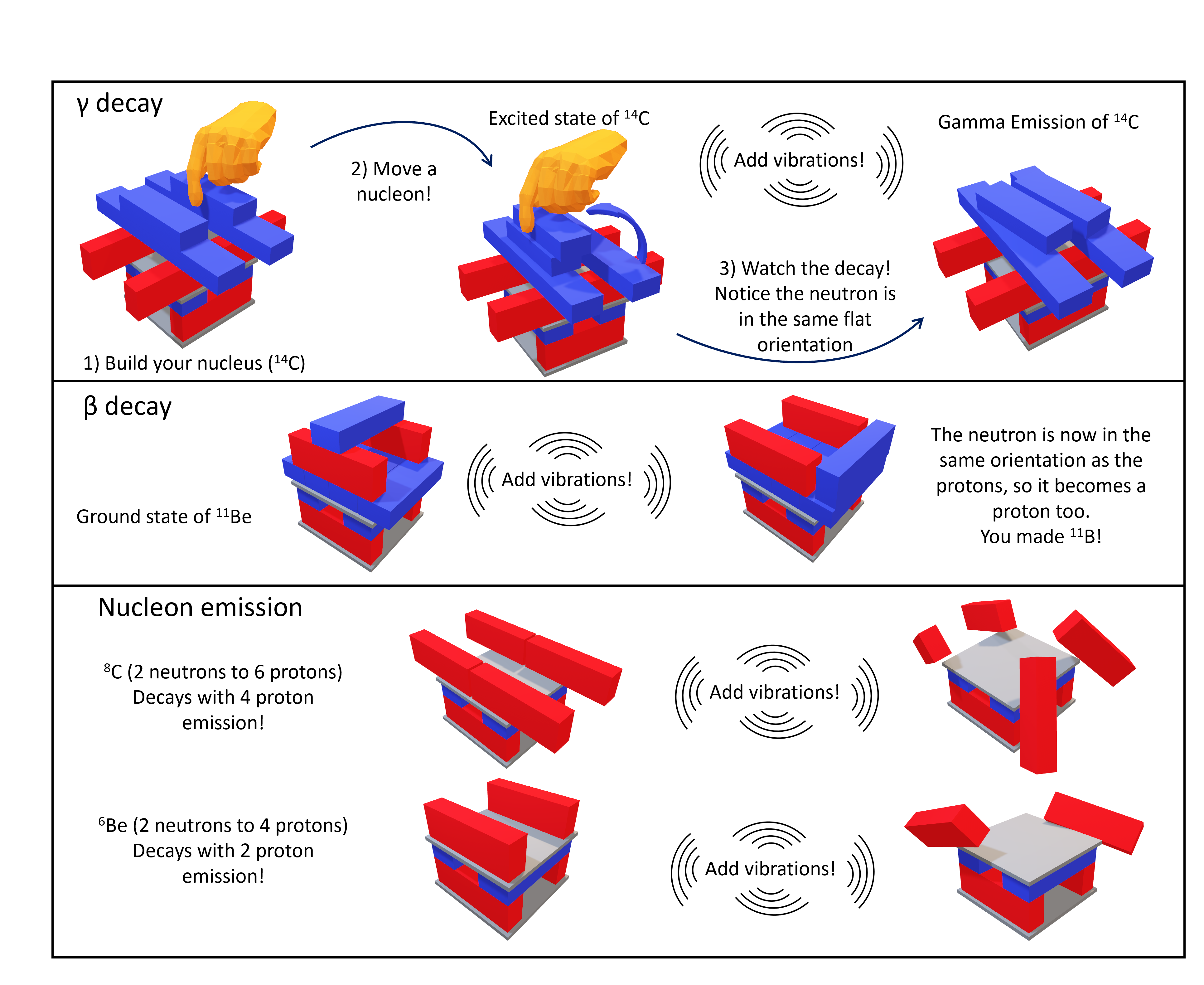}
    \caption{Types of decays observed in this demonstration. $\gamma$ decay involves a nucleon falling down to a lower level of the nucleus while retaining its original orientation. $\beta$ decay is when a nucleon falls down to a lower level of the nucleus or flips its orientation. Orientation change is defined as a transition between lying flat on its wide side to narrow side (neutron to proton) or vice versa. When nucleons fall off completely, we say that is nucleon emission.}
    \label{fig:decay_types}
\end{figure}

Finally, after a nucleus has been constructed, we can introduce vibrations to the table the nucleus sits on to simulate the effects of quantum mechanics.
As the table vibrates, some of the nucleons might fall apart or change orientation, representing various types of nuclear decays, see Figure~\ref{fig:decay_types}.
It is recommended to have something which vibrates at a consistent frequency and with different levels of strength to highlight motion of the blocks over time. Alternatively, participants can carefully shake or hit the table to vibrate the structure.
Some nuclei will survive the process, while others will inevitably decay in different and interesting ways.

\name has two possible goals: (1) build the largest, most stable, nucleus or (2) build the most exotic nucleus.
We define ``exotic" in light nuclei as a nucleus with an extreme proton-to-neutron ratio; for example, $^8$C or Carbon-8 containing 6 protons and only 2 neutrons (8 total nucleons) with a 3:1 ratio.
Generally, an exotic nucleus is one which has a short half-life, existing only for a few hours or less.
We have developed a companion website~\cite{Wylie2024} where users can explore the connections between \name and real nuclei, and one feature is to show the half-lives of nuclei that are exotic.
Additionally, we can make exotic nuclear states in a nucleus by using the same number of blocks, but stacked higher which represents an ``excited state" - see the two side branches B1 and B3 of the example chart in Figure \ref{fig:paper_rules} for an illustration of an excited $^3$He.

\subsection*{Competitive Play}\label{sec:competitive_play}

\begin{figure}[!h]
    \centering
    \includegraphics[width=\linewidth]{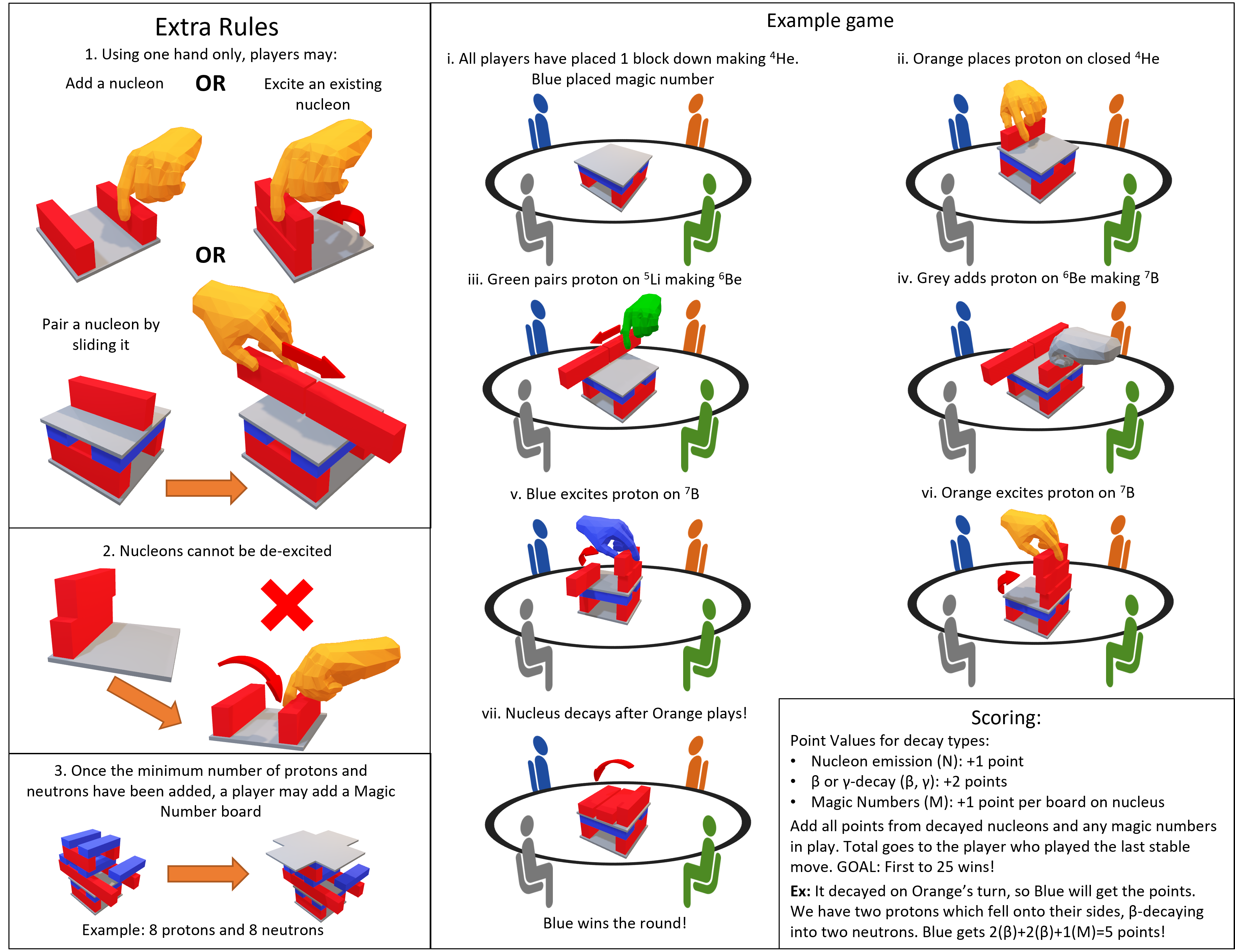}
    \caption{Competitive rules 1-3 are shown, with the lowest panel showing an example of how a round might be scored, with players taking turns clockwise around the table.}
    \label{fig:competitive_rules}
\end{figure}

For a more competitive demonstration, groups of participants can be formed and each take turns placing or moving blocks.
The goal of this mode is to have the group's nucleus fall apart on the next player's turn (after you have successfully played a move).
This is essentially the inverse process of other common block-building tower games participants may be familiar with.
We provide a point scale to encourage players to play moves which excite nucleons or create more exotic nuclei so there is more incentive to produce a variety of decay modes.
Furthermore, we encourage participants to play multiple rounds to explore the many ways nuclei can be made and decay.
Along with the basic rules, the following rules are included:
\begin{enumerate}
    \item Using one hand only, and not touching any other existing blocks, players may:
    \begin{enumerate}
        \item Place a new nucleon onto the nucleus
        \item Excite an existing nucleon by moving it to a level higher than its current position
    \end{enumerate}
    \item Nucleons cannot be de-excited (moved to a lower level)
    \item Once the minimum number of protons and neutrons have been added, a player may use their turn to place a Magic Number board instead of adding or exciting a nucleon
    \item (Optional) Nucleons cannot be added to levels below a Magic Number board unless it is to replace a vacancy left by an excited nucleon
\end{enumerate}
Specific examples of these rules and scoring are provided in Figure \ref{fig:competitive_rules}. The winner is the first player to reach 10 points. Points for each round are awarded to the player who made the last stable move, one that didn't result in a nuclear decay (e.g., when a nucleon falls off or down the nucleus). After the nucleus decays, add up the points for all decay types based on their assigned point values and any magic numbers in play. The point assignments are as follows:
\begin{itemize}
    \item Nucleon emission: when one or more nucleons fall completely off of the nucleus (1 point)
    \item $\beta$-decay: When a nucleon falls down to a lower level of the opposite nucleon type or onto the opposite side it started on i.e. a proton falls landing on side 2 changes to a neutron (2 points)
    \item $\gamma$-decay: When a nucleon falls down to a lower level of the same nucleon type or falls onto the same side it started on i.e. a proton falls landing on side 1 remains a proton (2 points)
    \item Additional points are added onto the decay score for each Magic Number board on the nucleus
\end{itemize}

The purpose of this version is to highlight how nucleons inside nuclei can be excited or oriented in unique configurations, leading to different decay modes. $\beta$ and $\gamma$ decays are very important to many phenomena including the formation of heavier elements in stars, so those decay modes are given higher point values. The primary takeaway is that these two decays do not change the total number of nucleons in a nucleus (in the context of the demonstration the blocks must remain on the nucleus) while nucleon emission reduces the total number of particles in a nucleus, as blocks fall off the nucleus onto the table.

Another aspect that players may come to appreciate is the complexity of nucleus formation. Rarely do nuclei form through straightforward, stable processes where protons or neutrons are added one at a time - like is done in the single-player version. In fact, outside effects (other players) lead to many exotic properties which might not normally be considered if playing in an isolated (single-player) view, as multiple players each compete against each other when building the nucleus.

\setcitestyle{numbers}
\bibliographystyle{unsrt} 